\documentclass[aps,floatfix,showpacs,showkeys,amsmath]{revtex4}
\usepackage{color,graphicx}
\usepackage{dcolumn}
\usepackage{bm}
\usepackage{epsfig}
\usepackage{supertabular}
\usepackage[bookmarksopen]{hyperref}
\def\be {\begin{equation}}
\def\ee {\end{equation}}
\def\nn {\nonumber}
\def\bea {\begin{eqnarray}}
\def\eea {\end{eqnarray}}

\begin{document} 
 \title{\Large {\bf Jet-dilepton conversion from an anisotropic quark-gluon plasma }}
\renewcommand{\thefootnote}{\alph{footnote}}
\bigskip
\bigskip
\author{Arghya Mukherjee}
\email{arghya.mukherjee@saha.ac.in}
\author{Mahatsab Mandal}
\email{mahatsab@gmail.com}
\author{ Pradip Roy}
\email{pradipk.roy@saha.ac.in}
\affiliation{Saha Institute of Nuclear Physics, 1/AF Bidhannagar
Kolkata - 700064, India}

\begin{abstract}
We calculate the yield of lepton pair production from jet-plasma interaction where the 
plasma is anisotropic in momentum space. We compare both the $M$ and $p_T$ distributions
from such process with the Drell-Yan  contribution. It is observed that the invariant 
mass distribution of lepton pair from such process dominate over the Drell-Yan up to
$3$ GeV at RHIC and up to $10$ GeV at LHC. Moreover, it is found that the contribution
from anistropic quark gluon plasma (AQGP) increases marginally compared 
to the isotropic QGP. In case of $p_T$-distribution we observe 
an increase by a factor of $3-4$ in the entire $p_T$-range at RHIC for AQGP. However, at LHC 
the change in the $p_T$-distribution is marginal as compared to the isotropic case. 
\end{abstract}
\maketitle

\section{Introduction}
The primary goal of heavy-ion collisions at Relativistic Heavy Ion Collider (RHIC) at BNL and 
Large Hadron Collider (LHC) at CERN is to establish the existence of a transient phase consisting
of quarks, anti-quarks and gluons known as Quark Gluon Plasma (QGP). Since such a phase 
lasts for a few $fm/c$, it is impossible to observe it directly. Thus, various indirect probes
have been proposed in the literature~\cite{prd44,zpc53,prc53,prd58,jhep0112,prd34,zpc46,prl70a,npa566,plb331,plb178,prc53a}. 
Electromagnetic probes is one of them. The advantage
of such probe is that once these are produced they can escape the interaction zone without much
distortion in their energy and momentum. They, thus carry the information of the collision dynamics 
very effectively. Photons and dileptons are produced throughout the evaluation process of the collisions.
In the low and intermediate mass region, lepton pair are produced from the $q\bar q\rightarrow l^+l^-$
process (thermal) and from various hadronic reactions and decay. In the high mass region, there is the contribution
from Drell-Yan process, which can be calculated from pQCD. Another important contribution in this invariant
mass region is the jet-dilepton conversion in the QGP. Several authors~\cite{prc67,prc74,npa865,prc92} 
have estimated this contribution
where a jet quark (anti-quark) interact with a thermal anti-quark (quark) to produce a large mass lepton
pair. It is to be noted that before annihilation of a $q~(\bar q)$ jet with a thermal $\bar q~(q)$, 
the $q~(\bar q)$ jet may lose energy. Such possibility has also been considered in Refs~\cite{prc74,npa865,prc92}. 
It has been observed in all those calculation that the magnitude of this mechanism is order of 
magnitude larger than the thermal processes and is of the same order of the Drell-Yan processes~\cite{prl25}.
It is to be noted that in all those calculations an isotropic plasma has been assumed to be formed.
But the most difficult problem lies in the determination of isotropization and thermalization
time scales ($\tau_{\rm iso}$ and $\tau_{\rm therm}$ ) of the QGP.
Studies on elliptic flow (upto about $p_T\sim 1.5−2$ GeV) using ideal 
hydrodynamics indicate that the matter 
produced in such collisions becomes isotropic with $\tau_{\rm iso}\sim 0.6$ fm/c~\cite{plb503}. 
On the contrary, perturbative estimates yield much slower
thermalization of QGP~\cite{plb502,prc71,jpg34}. However, recent hydrodynamical studies~\cite{prc78a}
have shown that due to the poor knowledge of the initial conditions, there is
a sizable amount of uncertainty in the estimation of thermalization  
or isotropization time. The other uncertain parameters
are the transition temperature $T_c$, the spatial profile, 
and the effects of flow. Thus it is necessary to find suitable 
probes which are sensitive to these parameters. As mentioned earlier
electromagnetic
probes have long been considered to be one of the most promising 
tools to characterize the initial state of the collisions~\cite{prd34,zpc46,prl70a,npa566,plb331}.
Dileptons (photon as well) can be one such observables.
In the early stages of heavy ion collisions, due to the rapid longitudinal 
expansion the plasma after formation in isotropic phase,
may become anisotropic~\cite{prd62,prd68,prd70,jhep08,prd70a, prl94,prd72,prd73,ahep}. 
As a result the momentum distribution of the plasma particles 
become anisotropic in momentum space. The author in Ref~\cite{prc78} have calculated the "medium"
dilepton yield for various isotropization times and compared it with Drell-Yan and jet-thermal processes.
It is shown that the effect of the anisotropy cannot be neglected while calculating $M-$ distribution
and $p_T-$  distribution. In fact in certain kinematic region this contribution is comparable to Drell-Yan
as well as jet-thermal process. Jet-photon conversion in the AQGP has been calculated in Ref~\cite{jpg37} with
$p_T$ up to $14$ GeV to extract the isotropization time. Also in intermediate and low $p_T$ the photon 
transverse momentum distribution has been calculated to infer about the isotropization time scale~\cite{prc79}. 
It is to be noted that the extracted values of $\tau_{\rm iso}$ from the above two cases are consistent. 
To the best of our knowledge, the contribution of the jet-dilepton conversion in AQGP has not been done so far. 
It is, thus, our purpose to estimate the dilepton yield from jet-plasma interaction in the present work. To keep the things 
simple in this work, we shall not include the energy loss of the jet in the AQGP. 

It should be noted that  in absence of any precise knowledge about the dynamics at early time of the collision,
one can introduce phenomenological models to describe the evolution of the pre-equilibrium
phase. In this work, we will use one such model, proposed in Ref.~\cite{prl100,prc78,prc84}, for the time 
dependence of the anisotropy parameter, $\xi(\tau)$, and hard momentum scale, $p_{\rm hard}(\tau)$. This model
introduces four parameters to parameterize the ignorance of pre-equilibrium dynamics: the
parton formation time ($\tau_i$), the isotropization time ($\tau_{\rm iso}$ ), which is the time when the system
starts to undergo ideal hydrodynamical expansion and $\gamma$ sets the sharpness of the transition
to hydrodynamical behavior. The fourth parameter $\delta$ is introduced to characterize the nature 
of pre-equilibrium anisotropy i.e. whether the pre-equilibrium phase is non-interacting
or collisionally broadened.

The plane of the paper is the following. In the next section we describe the formalism of jet-conversion dilepton
production in AQGP. Jet-production and Drell-Yan process will be discussed in section III. We present a brief discussion
on space-time evolution of AQGP in section IV. Section V will be devoted to present the results. Finally, we summarize in section VI.


\section{Formalism}
According to the relativistic kinetic theory, the dilepton production rate at leading
order in the coupling $\alpha$, is given by~\cite{plb283,prl70,plb331}:
\bea
\frac{dR^{l^+l^-}}{d^4P} = \int \frac{d^3{\bf p}_1}{(2\pi)^3}\frac{d^3{\bf p}_2}{(2\pi)^3}
f_{q/{\bar q}}({\bf p}_1)f_{\rm jet}({\bf p}_2)\,v_{12}\,\sigma^{l^+l^-}_{q{\bar q}}
\delta^{(4)}(P-p_1-p_2)\label{dirate}
\eea
where $f_{\rm jet}$ and $f_{q/{\bar q}}$ are the phase space distribution of the jet quarks/anti-quarks and medium 
quarks/anti-quarks respectively. The total cross section of the $q\bar{q}\rightarrow l^+l^-$ interaction 
is given by 
\be
\sigma^{l^+l^-}_{q{\bar q}} = \frac{4\pi}{3}\frac{\alpha^2}{M^2}N_cN_s
(1+\frac{2m_l^2}{M^2})(1-\frac{4m_l^2}{M^2})^{1/2}\sum_qe^2_q, 
\ee
where $N_c$ and $N_s$ are the color factor and spin factor, respectively. 
$m_l$ is the mass of lepton  and $M$ is the invarient mass of the lepton pair 
which is much greater than the 
dilepton mass. So we can easily ignore the lepton mass and we find 
$\sigma^{l^+l^-}_{q{\bar q}} = \frac{4\pi\alpha^2}{3M^2}N_cN_s\sum_qe^2_q$.
We also assume that the distribution function of quarks and anti-quarks is the same. 
$v_{12}$ is the relative velocity between the jet quark and medium quark/anti-quark:
\be
v_{12} = \frac{(p_1+p_2)^2}{2E_1E_2}.
\ee 

In this work we consider the medium is anisotropic in  momentum space so that the anisotropic distribution
function can be obtained  from an arbitrary isotropic distribution by squeezing or stretching along the 
preferred direction in the momentum space~\cite{prd68}:   
\be
f_{q/{\bar q}}({\bf p},\xi,p_{\rm hard}) = f_{q/{\bar q}}(\sqrt{{\bf p}^2+\xi(\tau)(\bf{ p}.\hat{ {\bf n}})^2},p_{\rm hard}(\tau)),
\ee
where $\hat{ \bf n}$ is the direction of anisotropy, $p_{\rm hard}(\tau)$ is a time-dependent hard momentum scale 
and $\xi(\tau)$ is a time-dependent parameter reflecting the strength of the momentum anisotropy. In isotropic case,
where $\xi = 0$, $p_{\rm hard}$ can be recognized with the plasma temperature $T$. 

The phase space distribution function for a jet, assuming the constant transverse density 
of the nucleus is given by~\cite{prl90,prc74}:
\bea
f_{\rm jet}({\bf p}) = \frac{(2\pi)^3}{g_{q}}\frac{{\mathcal{P}}(|\vec {\omega}_r|)}{\sqrt{\tau_i^2-z_0^2}}
\frac{1}{p_T}\frac{dN_{\rm jet}}{d^2p_Tdy}\delta(z_0)
\eea
where $g_q = 2\times 3$ is the spin and color degeneracy factor, 
$\tau_i\sim 1/p_T$ is the formation time of the quark or anti-quark jet, 
and $z_0$ is its position in the QGP
expansion direction. ${\mathcal{P}}(|\vec {\omega}_r|)$ is the initial 
jet production probability distribution at the radial position $\vec\omega_r$
in the plane $z_0$, where
\bea
|\vec\omega_r| &=& [\vec r-(\tau-\tau_i)\frac{\vec p}{|\vec p|}].\vec r\nn\\
&=& \sqrt{(r\cos\phi-\tau)^2+r^2\sin^2\phi}~~~{\rm for}~\tau_i\sim0
\eea
and  $\phi$ is the angle in the plane $z_0$ between the direction of the virtual
photon and the position where this virtual photon has been produced. 

Eq.~(\ref{dirate}) can be written as 
\bea
\frac{dR^{l^+l^-}}{d^4P} &=& \frac{5\alpha^2}{72\pi^5}\int \frac{d^3{\bf p}_1}{E_{p_1}}\frac{d^3{\bf p}_2}{E_{p_2}}
f_{q/{\bar q}}({\bf p}_1,\xi,p_{\rm hard})f_{\rm jet}({\bf p}_2)\delta^{(4)}(P-p_1-p_2)\nn\\
&=&\frac{5\alpha^2}{72\pi^5}\int \frac{d^3{\bf p}_1}{E_{p_1}E_{p_1}}
f_{q/{\bar q}}({\bf p}_1,\xi,p_{\rm hard})f_{\rm jet}({\bf p}-{\bf p}_1)\delta(E-E_{p_1}-E_{p_2})\Bigg|_{{\bf p}_2={\bf P}-{\bf p}_2}
\label{dirate1}
\eea
If we choose 
\bea
{\bf p}_1 &=& p_1(\sin\theta_{p_1}\cos\phi_{p_1},\sin\theta_{p_1}\sin\phi_{p_1},\cos\theta_{p_1}),\nn\\
{\bf P} &=& P(\sin\theta_{P}\cos\phi_{P},\sin\theta_{P}\sin\phi_{P},\cos\theta_{P})
\eea
and anisotropy vector $\hat{\bf n}$ along the $z$ direction, the $\delta$ function can be expressed as
\bea
\delta(E-E_{p_1}-E_{p_2}) = 2(E-p_1)\chi^{-1}\Theta(\chi)\sum_i^2\delta(\phi_i-\phi_{p_1}),
\eea
with $\chi=\Bigg[4P^2p_1^2\sin^2\theta_P\sin^2\theta_{p_1}-[2p_1(E-P\cos\theta_P\cos\theta_{p_1})-M^2]^2\Bigg]>0$ and 
the angle $\phi_i$ can be found by the solutions to the following equation:
\be
\cos(\phi_i-\phi_{p_1}) = \frac{2p_1(E-P\cos\theta_P\cos\theta_{p_1})-M^2}{2Pp_1\sin\theta_P\sin\theta_{p_1}}
\ee
Eq.~(\ref{dirate1}) can now be written as
\bea
\frac{dR^{l^+l^-}}{d^4P} &=& \frac{5\alpha^2}{18\pi^5}\int_{-1}^{+1}d(\cos\theta_{p_1})\int^{a_-}_{a_+}
\frac{dp_1}{\sqrt{\chi}}p_1f_{q/{\bar q}}(\sqrt{{\bf p}_1^2(1+\xi\cos^2\theta_{p_1})},p_{\rm hard})
f_{\rm jet}({\bf p}-{\bf p}_1)
\eea
with
\be
a_{\pm} = \frac{M^2}{2[E-P\cos(\theta_P\pm\theta_{p_1})]}
\ee

Now, the dilepton production rate $R$ is defined as the total number of lepton pair emitted from the 
4-dimensional space-time element $d^4x = \tau\, d\tau\, d\eta\, d^2x_{\perp}$ with $R=dN/d^4x$.
Here, $\tau=\sqrt{t^2-z^2}$ is the longitudinal proper time, $\eta=\tanh^{-1}(z/t)$ is the space-time 
rapidity, and ${\bf x}_{\perp}$ is a two-vector containing the transverse coordinates.

The total dilepton spectrum is given by a full space-time integration:
\bea
\frac{dN^{l^+l^-}}{dM^2dy} &=& \int_{p^{\rm min}_{T}}^{p^{\rm max}_{T}}d^2p_T\int_{\tau_i}^{\tau_f}\tau d\tau
\int_0^{R_{\perp}}rdr\int_0^{\pi}d\phi\int_{-\infty}^{+\infty} d\eta \frac{dR^{l^+l^-}}{d^4P}(E=m_T\cosh(y-\eta)),\\
\frac{dN^{l^+l^-}}{d^2p_Tdy} &=& \int_{M^{\rm min}}^{M^{\rm max}}dM^2\int_{\tau_i}^{\tau_f}\tau d\tau
\int_0^{R_{\perp}}rdr\int_0^{\pi}d\phi\int_{-\infty}^{+\infty} d\eta \frac{dR^{l^+l^-}}{d^4P}(E=m_T\cosh(y-\eta)),
\eea 
where $R_{\perp} = 1.2[<N_{\rm part}>/2]^{1/3}$ is the transverse dimension of the system amd $m_T$
is the transverse mass of the pair.
We have assumed that the plasma is formed at time $\tau_i$ and it undergoes a phase transition 
at transition temperature ($T_c$) which begins at the time $\tau_f$. $\tau_f$ is obtained by
using the condition $p_{\rm hard}(\tau = \tau_f) = T_c$. The energy of the dilepton 
pair in the fluid rest has to be understood as $E = m_T\cosh(y-\eta)$. 
Now the $\phi$ integration can be done as follows:
\be
\int_0^\pi d\phi\,{\mathcal P}({|\vec \omega_r|}) = 
\begin{cases}
0, & r^2+\tau^2-2\tau r\,>\,R^2_\perp \\
\frac{4}{R^2_\perp}(1-\frac{r^2+\tau^2}{R^2_\perp}), & r^2+\tau^2+2\tau r\,<\,R^2_\perp \\
\frac{4u_0}{\pi R^2_\perp}(1-\frac{r^2+\tau^2}{R^2_\perp})+\frac{8\tau r}{\pi R^4_\perp}\sin(u_0) &{\rm otherwise},
\end{cases}
\ee
where 
\be
u_0 = \arccos(\frac{r^2+\tau^2-R^2_\perp}{2\tau r}).
\ee

\section{Jets Production and Drell-Yan process}
The differential cross-section for the jet production in hadron-hadron collision ($A + B\rightarrow {\rm jets} + X$) 
can be written as~\cite{RMP59}
\bea
\frac{d\sigma_{\rm jet}}{d^2p_Tdy}=K\sum_{a,b}\int_{x_a^{\rm min}}^1 dx_a G_{a/A}(x_a,Q^2)G_{b/B}(x_b,Q^2)
\frac{x_ax_b}{x_a-\frac{p_T}{\sqrt{s}}e^y}\frac{1}{\pi}\frac{d{\hat\sigma}_{ab\rightarrow {\rm cd}}}{d{\hat t}} 
\eea
where $\sqrt{s}$ is the total energy in the center-of-mass and $x_a~(x_b)$ is the momentum fraction of the parton $a~(b)$ 
of the nucleon $A~(B)$. $G_{a/A}~(G_{b/B})$ is the parton distribution function (PDF) of the incoming parton $a~(b)$
in the incident hadron $A~(B)$. $K$ factor is used to account the next-to-leading (NLO) order effect. The minimum value of $x_a$ is 
\be
x^{\rm min}_a = \frac{p_T e^y}{\sqrt{s}-p_T e^{-y}}.
\ee
The value of the momentum fraction $x_b$ can be written as
\be
x_b = \frac{x_a p_T e^{-y}}{x_a \sqrt{s}-p_T e^y}. 
\ee
$\frac{d{\hat\sigma}_{ab\rightarrow {\rm cd}}}{d{\hat t}}$ is the cross section of parton collision 
at leading order. These process are: $qq\rightarrow qq$, $q\bar q\rightarrow q \bar q$, 
$q\bar q\rightarrow q^\prime\bar q^\prime$, $qq^\prime\rightarrow qq^\prime$,
$q\bar q^\prime\rightarrow q\bar q^\prime$, $qg\rightarrow qg$, and 
$gg\rightarrow q\bar q$. The yield for producing jets in the heavy-ion collision
is given by
\be
\frac{dN_{\rm jet}}{d^2p_Tdy} = T_{\rm AA} \frac{d\sigma_{\rm jet}}{d^2p_Tdy}\Bigg|_{y=0}\label{yield2}
\ee
where $T_{\rm AA} = 9A^2/8\pi R^2_\perp$  is the nuclear thickness function for zero impact parameter~\cite{prl90}.
The $p_T$ distribution of the jet quarks in the central rapidity region ($y=0$) was computed in~\cite{prc67} 
and parameterized as 
\be
\frac{dN_{\rm jet}}{d^2p_T dy}{\Bigg |}_{y=0} = K \frac{a}{(1+\frac{p_T}{b})^c}\label{yield1}.
\ee
Numerical values for the parameters $a$, $b$ and $c$ are listed in Ref.\cite{prc67}.   

The cross-section for the Drell-Yan process (LO) is given by~\cite{plb78}
\be
\frac{dN_{\rm DY}}{dM^2dy}=T_{\rm AA} K_{\rm DY} \frac{4\pi\alpha}{9M^4}
\sum_q e^2_q \big[x_1 G_{q/A}(x_1,Q^2)x_2 G_{\bar q/B}(x_2,Q^2)
+\sum_q e^2_q \big[x_1 G_{\bar q/A}(x_1,Q^2)x_2 G_{q/B}(x_2,Q^2)\big]
\ee
where the momentum fractions with rapidity $y$ are
$x_1= \frac{M}{\sqrt s}e^y$, $x_2= \frac{M}{\sqrt s}e^{-y}$. 
$K_{\rm DY}$ factor of 1.5 is used to account for the NLO correction~\cite{plb473}. 
\section{Space-time evaluation}
For the case of expanding plasma, we will be required to specify a proper-time dependence
of the anisotropy parameter, $\xi$ and the hard momentum scale, $p_{\rm hard}$.
In our calculation, we assume an isotropic plasma is formed at initial time $\tau_i$ and 
initial temperature $T_i$. The initial rapid expansion of the matter along the longitudinal 
direction causes faster cooling in this direction than in the transverse direction~\cite{plb502} 
and as result, a local momentum-space anisotropy occurs and remains until $\tau = \tau_{\rm iso}$. 
In this work, we shall follow the work of ref~\cite{prl100,prc78} to evaluate the dilepton production
rate from the first few Fermi of the plasma evolution. According to this model there can be 
three possible scenarios : 
(i) $\tau_i=\tau_{\rm iso}$, the system evolves hydrodynamically 
so that $\xi(\tau) = 0$ and we can identify the hard momentum scale with the plasma temperature so that 
$p_{\rm hard}(\tau) = T(\tau) = T_0(\tau_i/\tau)^{1/3}$,  
(ii)$\tau_{\rm iso}\rightarrow \infty$, the system never comes to equilibrium, 
(iii)$\tau_{iso}\geq\tau_i$ and $\tau_{\rm iso}$ is finite, one should  devise a time evolution 
model for $\xi(\tau)$ and $p_{\rm hard}(\tau)$ which smoothly interpolates between pre-equilibrium 
anisotropy and hydrodynamics and we shall follow scenario (iii). The time dependent parameters 
($\xi, p_{\rm hard}$), are obtained in terms of a smeared step function~\cite{prl100}: 
\be
\lambda(\tau)=\frac{1}{2}(\tanh[\gamma(\tau-\tau_{iso})/\tau_i]+1). 
\label{eq_lamda}
\ee
where the transition width, $\gamma^{-1}$ is introduced to take into account the smooth transition between
non-equilibrium and hydrodynamical evolution at $\tau = \tau_{\rm iso}$. It is clearly seen that for 
$\tau\,<<\,\tau_{\rm iso}$, we have $\lambda\rightarrow0$, corresponding to anisotropic evolution and for 
$\tau\,>>\,\tau_{\rm iso}$, $\lambda\rightarrow1$ which corresponds to hydrodynamical evolution.

With this, the time dependence of relevant quantities are as follows~\cite{prl100,prc78}:
\bea
p_{\rm hard}(\tau)&=&T_i\,\left[{\cal U(\tau)}/{\cal U}(\tau_i)\right]^{1/3},\nn\\
\xi(\tau) &=& a^{\delta[1-\lambda(\tau)]}-1,
\label{eq_xirho}
\eea
where ${\mathcal U}(\tau)\equiv\left[{\mathcal R}
\left(a_{\rm iso}^\delta-1\right)\right]^{3\lambda(\tau)/4}
\left(a_{\rm iso}/a\right)^{1-\delta[1-\lambda(\tau)]/2}$, 
$a\equiv \tau/\tau_i$ and $a_{\rm iso}\equiv\tau_{\rm iso}/\tau_i$ and 
${\mathcal R}(x) = \frac{1}{2}(\frac{1}{1+\xi}+\frac{\arctan\sqrt{\xi}}{\sqrt{\xi}}). $
In the present work, we have used a {\em free streaming interpolating} model 
that interpolates between early-time $1+1$ dimensional longitudinal free streaming and late-time  
$1+1$ dimensional ideal hydrodynamic expansion by choosing $\delta=2$.

As the colliding nuclei do have a transverse density profile,
we assume that the initial temperature profile is given by~\cite{prc72}
\begin{eqnarray}
T_i(r) = T_i\,\left[2\left(1-r^2/R_A^2\right)\right]^{1/4}
\label{eq16}
\end{eqnarray}
Using Eqs.(\ref{eq_xirho}) and (\ref{eq16}) we obtain the profile of the hard momentum scale as
\begin{eqnarray}
 p_{\rm hard}(\tau,r) = T_i\,\left[2\left(1-r^2/R_A^2\right)\right]^{1/4} 
{\bar {\cal U}}^{c_s^2}(\tau)\label{eq17}.
\end{eqnarray}

In case of isentropic expansion the initial temperature ($T_i$) and thermalization time ($\tau_i$)
can be related to the observed particle rapidity density by the following equation~\cite{prd32}:
\be
T_i^3(b_m)\tau_i = \frac{2\pi^4}{45\zeta(3)} \frac{1}{\pi R_{\perp}} \frac{1}{4 a_k}<\frac{dN}{dy}(b_m)>,
\ee
where $\frac{dN}{dy}(b_m)$ is the hadron multiplicity for a given centrality class with maximum impact
parameter $b_m$, $R_\perp$ is the transverse dimension of the system,
$\zeta(3)$ is the Riemann zeta function, and $a_k = (\pi^2 /90) g_k$ for a plasma of 
massless u, d and s quarks and gluons, where $g_k = 42.25$.

\section{Results}
For central Au $+$ Au collision at $\sqrt {S_{NN}} = 200$ GeV we fix our initial conditions to be 
$T_i = 446$ MeV and $\tau_i = 0.147$ fm/c for the plasma phase. For the LHC at  $\sqrt{S_{NN}} =5.5$ TeV,
our initial conditions are $T_i = 845$ MeV and $\tau_i = 0.088$ fm/c.  
\begin{figure}[htb]
\begin{center}
\includegraphics[height=5cm,,angle=0]{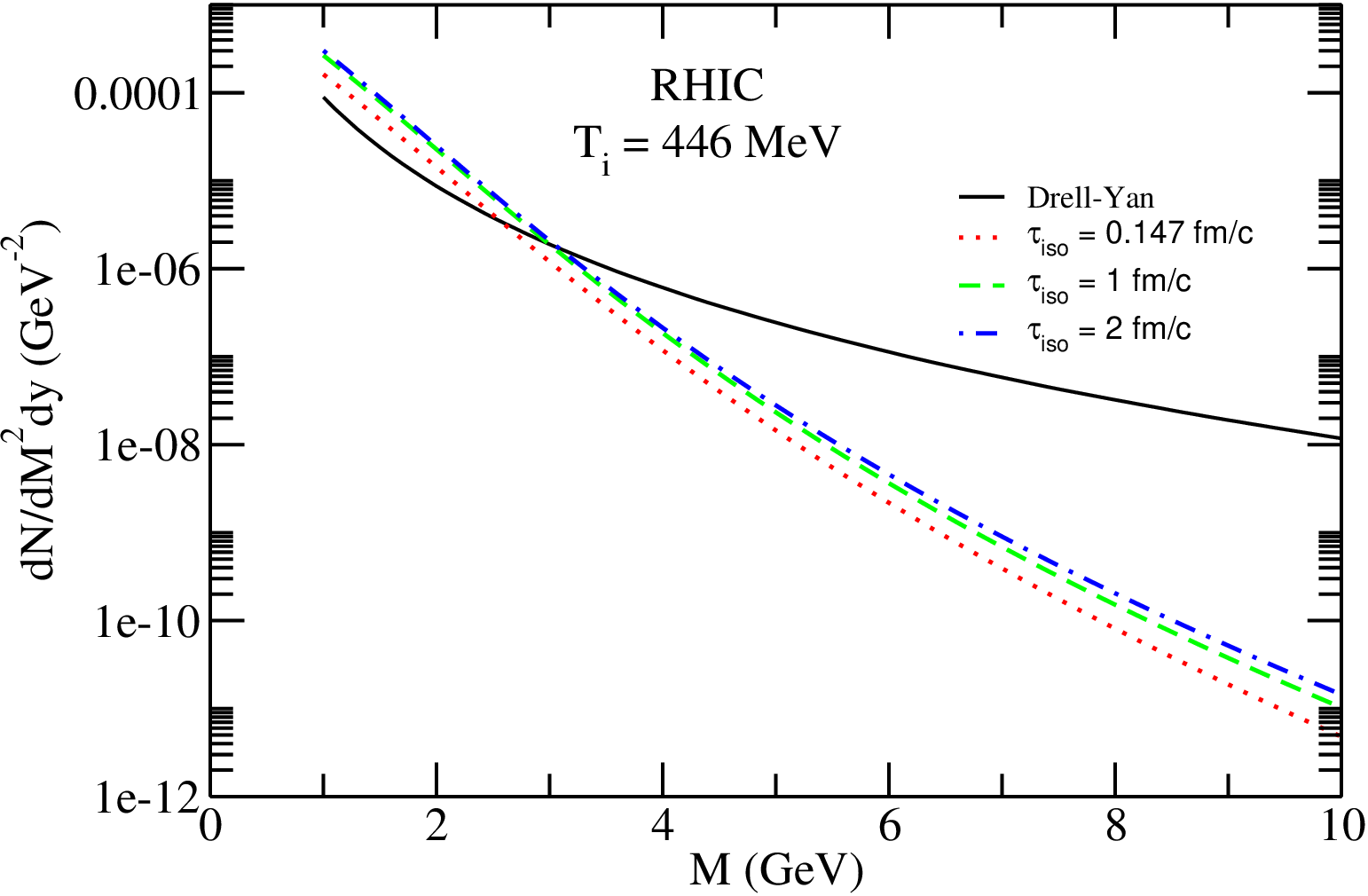}~~~
\includegraphics[height=5cm,,angle=0]{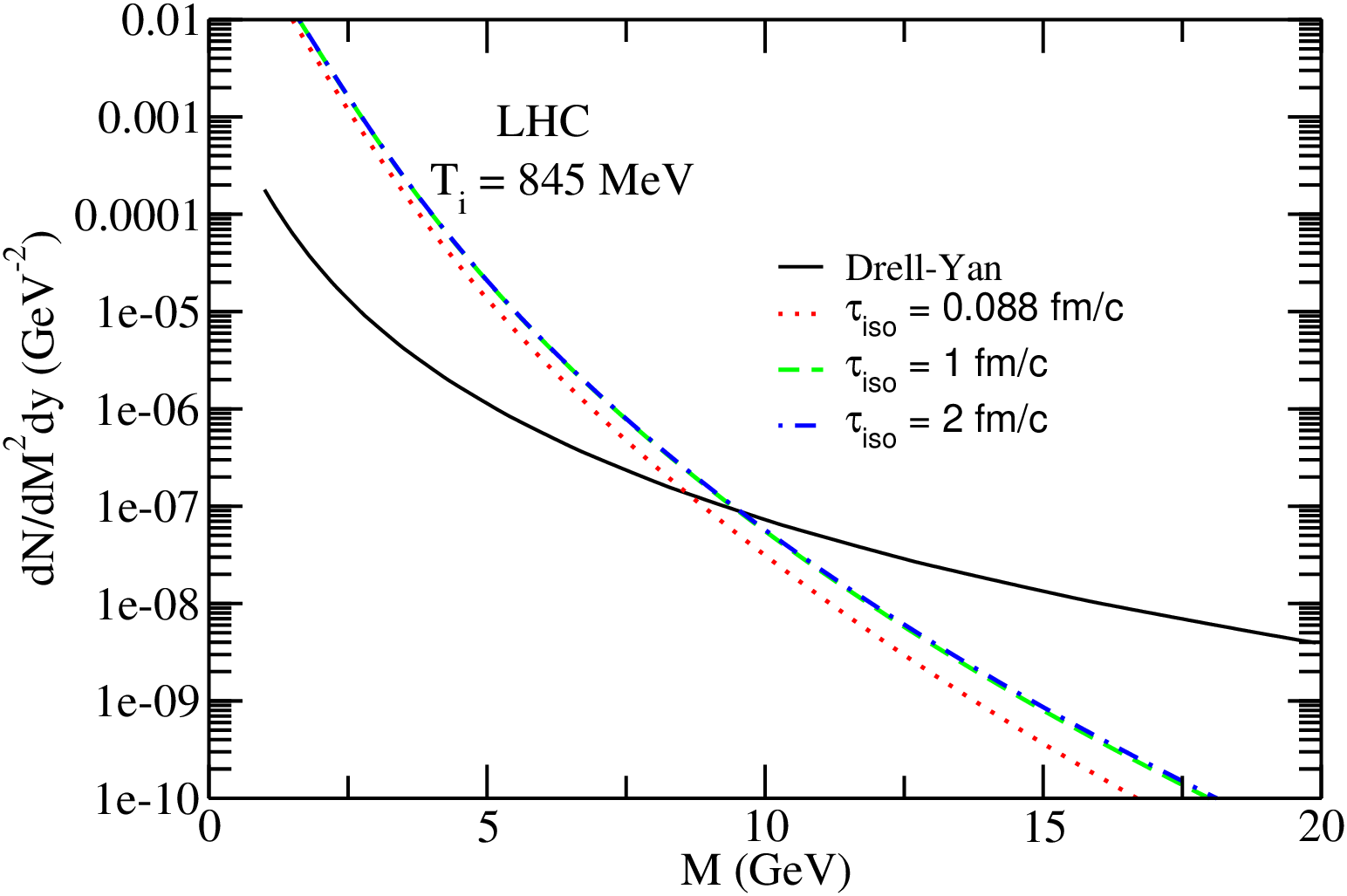}
\caption{Dilepton yield for central Au $+$ Au collisions at $\sqrt{S_{NN}} = 200$ GeV (left panel) and 
for central Pb $+$ Pb collisions at $\sqrt{S_{NN}} = 5.5$ TeV (right panel). }
\label{fig1}
\end{center}
\end{figure}

\begin{figure}[htb]
\begin{center}
\includegraphics[height=5cm,,angle=0]{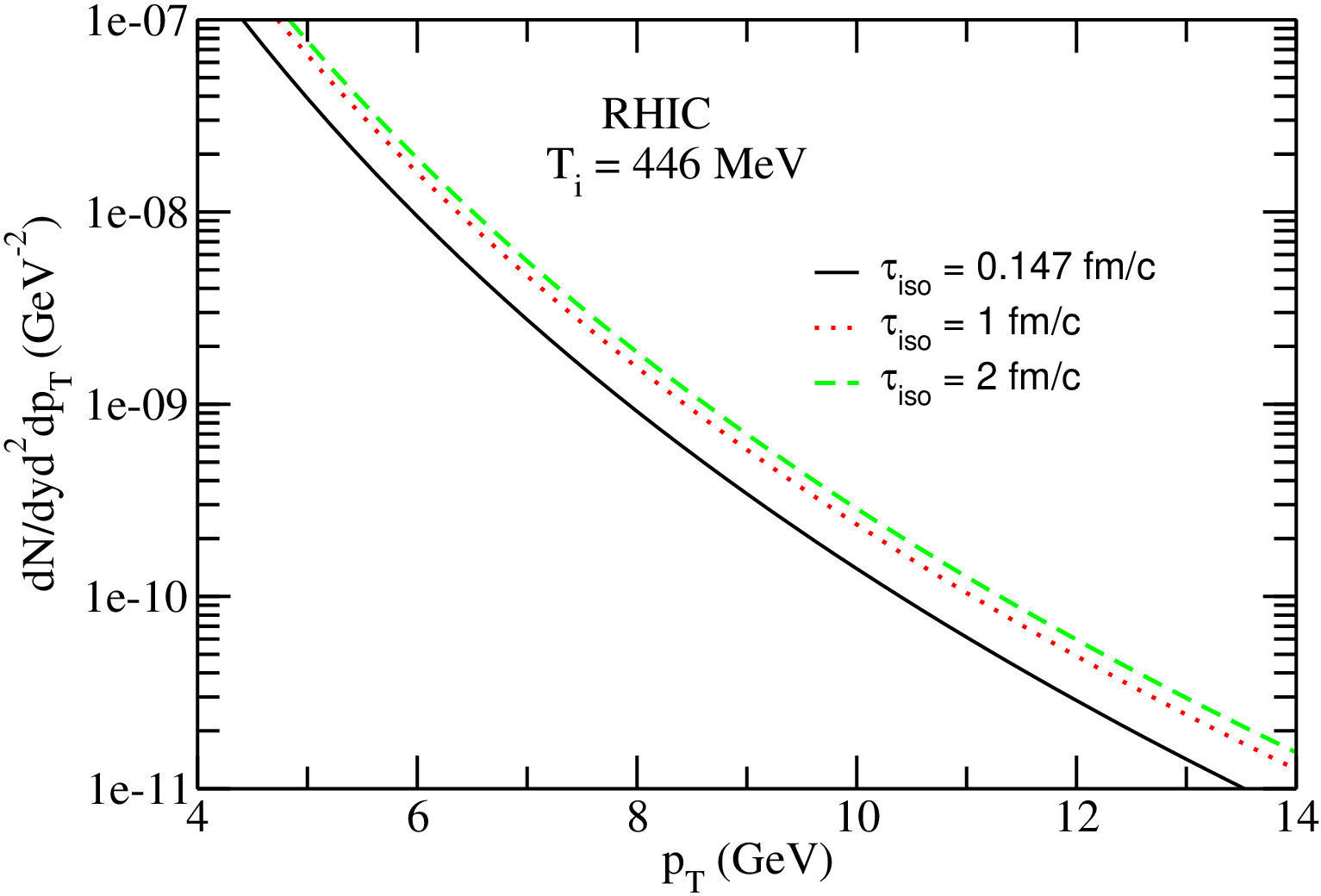}~~~
\includegraphics[height=5cm,,angle=0]{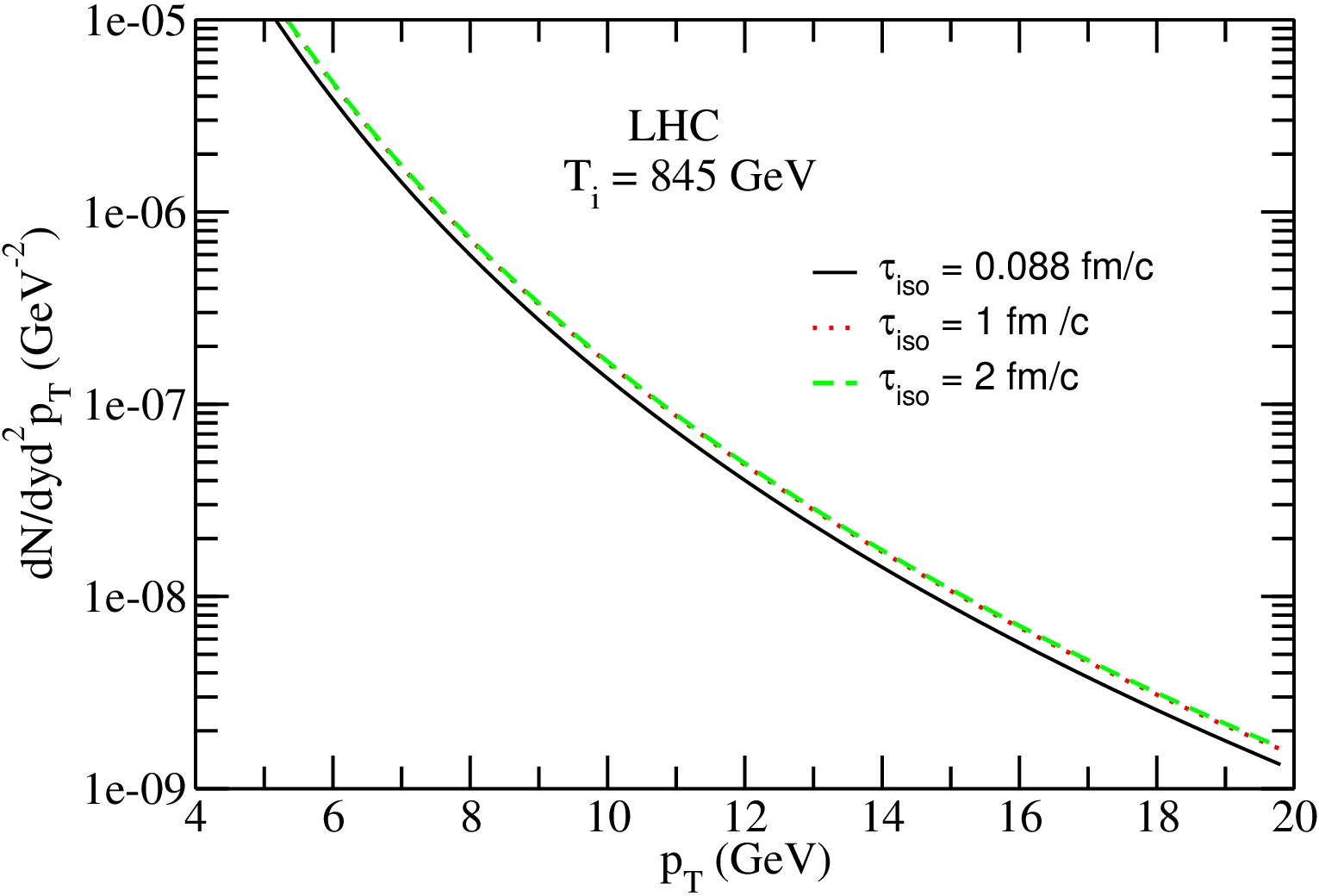}
\caption{$p_T$ distribution of the jet-conversion dilepton, integrated in the range $0.5\,<\,M\,<1$ GeV, 
for the RHIC (left panel) and the LHC (right panel). }
\label{fig2}
\end{center}
\end{figure}

\begin{figure}[htb]
\begin{center}
\includegraphics[height=5cm,,angle=0]{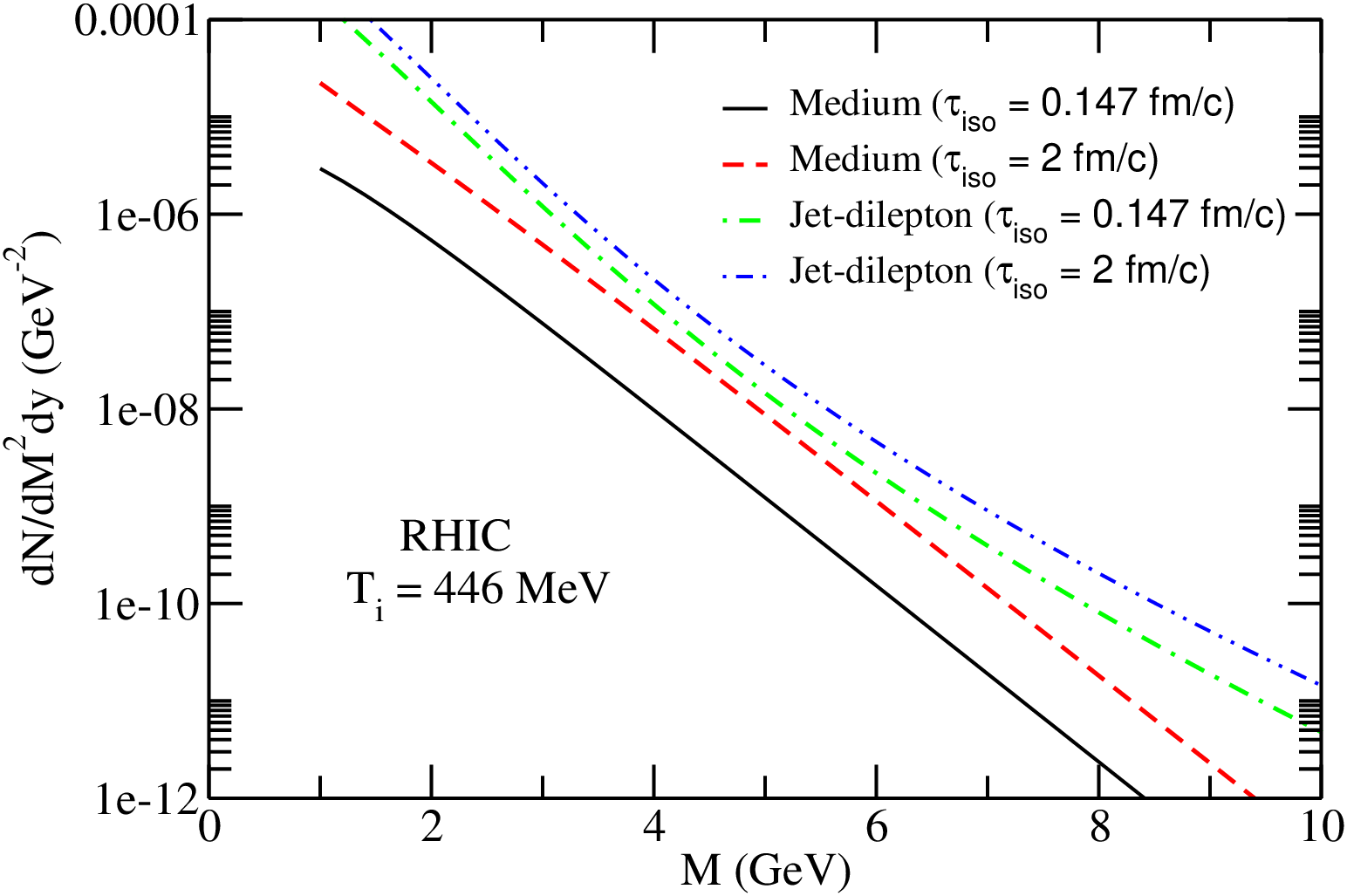}~~~
\includegraphics[height=5cm,,angle=0]{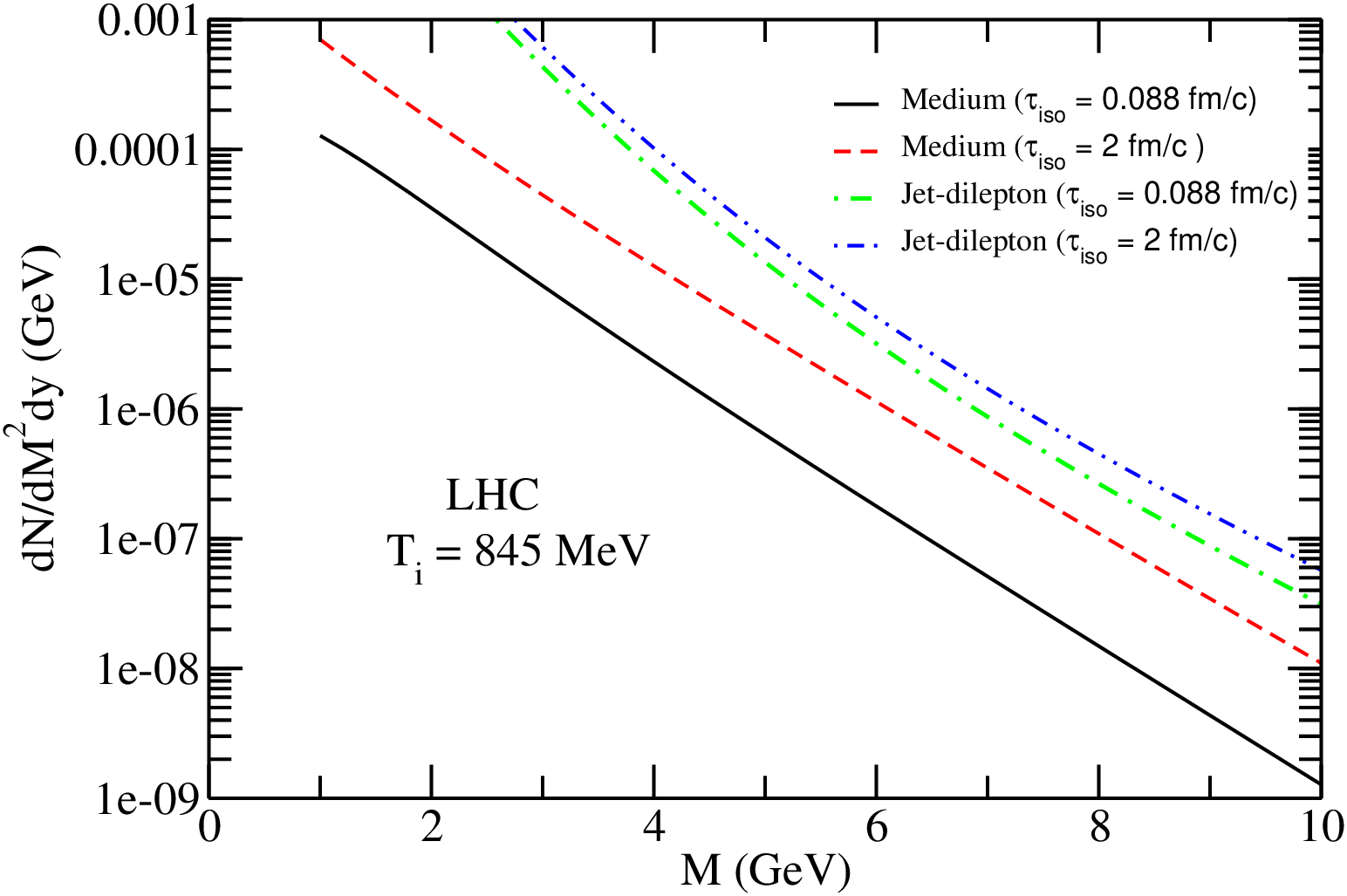}
\caption{Collisionally broadened interpolating model dilepton yield and jet-conversion dilepton yield as a function
of invariant mass in central Au $+$ Au collisions at $\sqrt{S_{NN}} = 200$ GeV (left panel) and for central 
Pb $+$ Pb collisions at $\sqrt{S_{NN}} = 5.5$ TeV (right panel). }
\label{fig3}
\end{center}
\end{figure}

Using the above initial conditions, we display the invariant mass distribution
for RHIC (left panel) and LHC (right panel) energies in Fig.~(\ref{fig1}).
It is found that the contribution from jet-dilepton conversion in
isotropic plasma  dominates over the Drell-Yan contribution up to $M \sim
2.5 $ GeV (see left panel). However, when anisotropy is introduced this
threshold increases to $M = 3$ GeV irrespective of the values of the
isotropization time. For LHC energies (right panel) this threshold increases
up to $M\sim 10 $ GeV providing an expanded window at LHC where jet-conversion
dilepton could be observed when anisotropy is taken into account.
The $p_T$-distribution is shown in Fig.~(\ref{fig2}) where we find that 
the effect of anisotropy at RHIC energies is substantial (increases by a factor 
of $4$). Surprisingly, at LHC energies the effect is not substantial. In Fig.~(\ref{fig3}) 
we compare the contribution from jet-dilepton conversion with the medium dilepton~\cite{prc78},
where $f_{\rm jet}(\bf {p_2})$ in Eq.~(\ref{dirate}) is replaced by $f_{\bar q/q}({\bf p_2})$.
It is found that both for the RHIC and LHC energies at various $\tau_{\rm iso}$, the medium
dilepton always remains below the jet-dilepton conversion in AQGP. 

\section{Summary and Discussion}
  In this work we have calculated the contribution of lepton pair production
in high mass region from jet-plasma interactions in AQGP. For the simplicity,
(1+1)d anisotropic hydrodynamics has been used as the effect of transverse
expansion will be marginal in the early stage of the collision (note that
momentum-space anisotropy is an early time phenomenon). 
It is found that the threshold value of $M$ beyond which DY process dominates
over jet-conversion dilepton increases marginally with the introduction of momentum-space
anisotropy both for RHIC and LHC. However, we do not find  any appreciable
change in the $p_T$ distribution at LHC energies even if the anisotropy is introduced.
We have also shows that the medium dilepton contribution always remains below the dilepton
from the jet plasma interaction. In fact, the former is less than by a factor $4-10$
depending upon $\tau_{\rm iso}$ and $p_T$. Finally, in the present calculation
we have not included the energy loss of the energetic jets 
which we propose to report elsewhere~\cite{AM}. We are also in the process
of applying (1+3)d anisotropic hydrodynamics to estimate the jet-conversion
dilepton from anisotropic quark-gluon plasma.


\begin{thebibliography}{99}
\bibitem{prd44} J. Kapusta, P. Lichard and D. Seibert, Phys. Rev. D {\bf 44}, 2774 (1991), Erratum ibid {\bf 47}, 4171 (1993).
\bibitem{zpc53} R. Baier, H. Nakkagawa, A. Niegawa and  K. Redlich, Z. Phys. C {\bf 53}, 433 (1992).
\bibitem{prc53} P. K. Roy, D. Pal, S. Sarkar, D. K. Srivastava, and B. Sinha, Phys. Rev. C {\bf 53} 2364 (1996).
\bibitem{prd58} P. Aurenche, F. Gelis, H. Zaraket, and R. Kobes, Phys. Rev. D {\bf 58}, 085003 (1998).
\bibitem{jhep0112} P. Arnold, G. D. Moore and L. G. Yaffe, JHEP 0112, 009 (2001).
\bibitem{prd34} K. Kajantie, J. Kapusta, L. Mclerran, A. Mekjian, Phys. Rev. D {\bf 34}, 2746 (1986).
\bibitem{zpc46} K. J. Eskola and J. Lindfors, Z. Phys. C {\bf 46}, 141 (1990).
\bibitem{prl70a} K. Geiger and J.~I.~Kapusta, Phys. Rev. Lett. 70 1920 (1993).
\bibitem{npa566} B. Kampfer and O. P. Pavlenko, Nucl. Phys. A {\bf 566}, 351 (1994).
\bibitem{plb331} M.~Strickland Phys. Lett. B 331 (1994) 245.
\bibitem{plb178} H. Satz and T. Matsui, Phys. Lett. {\bf B178}, 416 (1986).
\bibitem{prc53a} X. M. Xu, D. Kharzeev, H. Satz, and X. N. Wang, Phys. Rev. C {\bf 53}, 3051 (1996).

\bibitem{prc67} D.~K.~Srivastava, C.~Gale, R.~J.~Fries, Phys. Rev. C 67  034903 (2003).
\bibitem{prc74} S.~Turbide, C.~Gale,  D.~K.~Srivastava, R.~J.~Fries, Phys. Rev. C 74  014903 (2006).
\bibitem{npa865} Yong-Ping Fu and Yun-De Li, Nucl. Phys. A {\bf 865} 76 (2011).
\bibitem{prc92} Yong-Ping Fu and Q. Xi Phys. Rev. C {\bf 92}, 024914 (2015).

\bibitem{prl25} S. D. Drell and T.  M. Yan, Phys. Rev. Lett. {\bf 25}, 316 (1970).

\bibitem{plb503} P. Huovinen, P. F. Kolb, U. W. Heinz, P. V. Ruuskanen and S. A. Voloshin, Phys. Lett. {\bf B503}, 58 (2001).

\bibitem{plb502} R. Baier, A. H. Mueller, D. Schiff and D. T. Son, Phys. Lett. {\bf B502}, 51 (2001).
\bibitem{prc71} Z. Xu and C. Greiner, Phys. Rev. C {\bf 71}, 064901 (2005).
\bibitem{jpg34} M. Strickland, J. Phys. G {\bf 34}, S429 (2007).

\bibitem{prc78a} M. Luzum and P. Romatschke Phys. Rev. C {\bf 78}, 034915 (2008), Erratum ibid {\bf 79}, 039903 (2009).


\bibitem{prd62} S. Mrowczynski and M. H. Thoma, Phys. Rev. D {\bf 62}, 036011 (2000).
\bibitem{prd68} P.~Romatschke and M.~Strickland, Phys. Rev. D {\bf 68} 036004 (2003). 
\bibitem{prd70}  P.~Romatschke and M.~Strickland, Phys. Rev. D {\bf 70} 116006 (2004).
\bibitem{jhep08} P. Arnold, J. Lenaghan, and G. D. Moore, JHEP 08, 002 (2003).
\bibitem{prd70a} S. Mrowczynski, A. Rebhan and M. Strickland, Phys. Rev. D {\bf 70}, 025004 (2004).
\bibitem{prl94} A. Rebhan, P. Romatschke and M. Strickland, Phys. Rev. Lett. {\bf 94}, 102303 (2005).
\bibitem{prd72} P. Arnold, G. D. Moore and L. G. Yaffe, Phys. Rev. D {\bf 72},054003 (2005).
\bibitem{prd73} B. Schenke,  M. Strickland, C. Greiner and M. H. Thoma, Phys. Rev. D {\bf 73}, 125004 (2006). 
\bibitem{ahep} M. Mandal and P. Roy, Adv. High Energy Phys. 2013,  371908 (2013).

\bibitem{prc78} M. Martinez and M. Strickland, Phys. Rev. C {\bf  78}, 034917 (2008).

\bibitem{jpg37} L. Bhattacharya and P. Roy, J. Phys. G {\bf 37}, 105010 (2010). 

\bibitem{prc79} L. Bhattacharya and P. Roy, Phys.Rev.C {\bf 79}, 054910 (2009).

\bibitem{prl100} M. Martinez and M. Strickland, Phys. Rev. Lett. 100, 102301 (2008).
\bibitem{prc84}M. Mandal, L. Bhattacharya and P. Roy, Phys. Rev. C {\bf 84} 044910 (2011).

\bibitem{plb283} J.~I.~Kapusta, L.~D.~McLerran and D.~K.~Srivastava,  Phys. Lett. B 283 145 (1992).
\bibitem{prl70} A.~Dumitru, D.~H.~Rischke, T.~Schonfeld, L.~Winckelmann, H.~Stocker and W.~Greiner  Phys. Rev. Lett. 70 2860 (1993).


\bibitem{prl90} R.~J.~Fries, B.~Muller, D.~K.~Srivastava, Phys. Rev. Lett. 90 132301 (2003).
 
\bibitem{RMP59} J.~F.~Owens Rev.  Mod. Phys. 59 465 (1987). 

\bibitem{plb78} E.~V.~Shuryak, Phys. Lett. B 78 150 (1978).
\bibitem{plb473} R.~Rapp, E.~V.~Shuryak, Phys. Lett. B 473 13 (2000).   
  
 


\bibitem{prc72} S. turbide, C. Gale, S. Jeon, and G. D. Moore , Phys. Rev. C {\bf 72}, 014906 (2005).
\bibitem{prd32} R. C. Hwa and K. Kajantie, Phys. Rev. D {\bf 32}, 1109 (1985).

\bibitem{AM} A. Mukherjee et al 2016 in preparation. 

\end{thebibliography}
\end{document}